\documentclass[12pt]{article}
\usepackage{amssymb,graphicx}

\newcommand{\eqref}[1]{(\ref{#1})}

\newcommand{\be}{\begin{equation}}
\newcommand{\ee}{\end{equation}}

\newcommand{\bea}{\begin{eqnarray}}
\newcommand{\eea}{\end{eqnarray}}

\def\x{{\bf x}}

\def\lag{\langle}
\def\rag{\rangle}

\def\removetext#1{{}}

\begin{document}
\title{{Finite temperature spectral functions}\\ {of the linear $O(N)$-model 
at large $N$}\\ {applied to the $\pi -\sigma$ system}}
\author{
A. Patk{\'o}s$^{a }$\footnote{Dept. of Atomic Physics,
e-mail: patkos@ludens.elte.hu}, Zs. Sz{\'e}p$^{a
}$\footnote{ Dept. of Atomic Physics, 
e-mail: szepzs@cleopatra.elte.hu}
~and~P. Sz{\'e}pfalusy$^{a,b }$\footnote{Dept. of Physics of Complex Systems, 
e-mail: psz@complex.elte.hu}
\\ $^{a}$E{\"o}tv{\"o}s University, H-1117 Budapest,
Hungary\\$^{b}$
Research Institute for Solid State Physics and Optics,\\
 H-1525, Budapest, Hungary}

\maketitle
\begin{abstract}
The thermal evolution of the spectral densities derivable from the 
two-point functions of the elementary  and the quadratic composite fields
 of the $O(N)$ model is studied in the isosinglet channel and in the broken
symmetry phase at infinite $N$. The results are applied
with realistic parameter values to the $N=4$ case. They provide
a reasonable description of the $\sigma$ meson at $T=0$. Threshold
enhancement is observed around $T\sim 1.07m_\pi$. For higher
temperatures the maximum of the  spectral function in the
single meson channel decreases and becomes increasingly
rounded. 
\end{abstract}
\section{Introduction}
There is increasing interest in finding signatures for the characterisation
of strongly interacting matter around the transition temperature from the
hadronic to the quark-gluon regime. Density and temperature induced
features of the 
correlations in the soft pion spectra is a frequently discussed issue. 
The majority of these pions comes from resonance decay. One expects that
temperature induced shifts in the position and width of the $\rho$ and 
$\sigma$ pole location is sensitively reflected in the spectra. 

The effects related to the variation of the location and the width of
the broad 
$\sigma$ resonance were investigated in the past decade systematically
\cite{hatsuda01}. The main theoretical framework of these investigations was 
the $O(4)$ linear sigma-model. Various reorganisations of the perturbative
series \cite{chiku98} were proposed for the self-energy 
calculations.  Despite of the rather large value of the nonlinear 
coupling $\lambda_R\approx{\cal O}(30-70)$, 
one loop calculations gave important qualitative hints for the 
enhancement of the spectral function in the $\sigma$-channel  near the 
two-pion threshold as the temperature approaches the transition region.

In the non-relativistic context it has been recognised long time ago that
  in scalar theories the hybridisation 
\cite{kondor74} of the two-point functions of the elementary and
quadratic composite fields play essential 
role in the determination of the elementary excitations 
in the whole temperature interval of the broken symmetry phase. 
Neutron scattering data on superfluid helium were successfully interpreted by
taking into account the hybridisation\cite{griffin93}.
   
Large $N$ techniques are known to account for this phenomenon
correctly, while ordinary perturbative approaches usually miss it
\cite{kondor74,szepfalusy76}.

An additional point in favour of the use of the large $N$ expansion
is that it leads to a second order transition point in agreement
 with  the well-known behaviour of the $O(N)$ model\cite{cooper97}. The
lowest order computations done with any 
(even optimised) version of the conventional perturbation theory yield a 
first order phase transition when the temperature is increased. 

Large $N$ techniques are extensively studied also
in connection with the out-of-equilibrium evolution and thermalisation 
of relativistic Bose-condensates
\cite{cooper94,boyan99,baacke00,berges01,aarts02}, though the
hybridisation in this respect did not receive till now much attention.

Our main goal with this investigation was to apply large $N$ techniques for
finding the finite temperature variation of the spectral function
in the elementary and quadratic composite isosinglet channels, which 
influences the $\pi -\pi$ 
correlations measured in heavy ion collisions to a large extent. For this
purpose the parametrisation of the linear $O(4)$-symmetric meson model is 
chosen to reproduce as
closely as possible the accepted zero temperature fit to the complex mass,
$M_\sigma -i\Gamma /2$ of this broad resonance \cite{tornquist02}.

In this Letter, after summarising the general formalism,
mostly the results of the chirally invariant limiting case will be discussed, 
where the pions are massless, since the main ideas 
can be presented in this case very transparently. Results relevant for the
case with explicit symmetry breaking will be presented shortly in the 
concluding part of the paper.
 
We shall work in the leading order large $N$ approximation
to the Schwinger-Dyson equations of the $O(4)$ model in the broken symmetry
phase. The effect of next-to-leading order corrections is the subject
of our ongoing research. 
        
\section{The model and its renormalised dynamical equations}

The Lagrangian density including a term reflecting explicit breaking of the
$O(N)$ symmetry is the following:
\be
L={1 \over
2}[\partial_\mu\phi^a\partial^\mu\phi^a-m^2\phi^a\phi^a]-{\lambda\over
24N} (\phi^a)^2(\phi^b)^2+\sqrt{N}h\phi^1.
\ee
The broken symmetry phase can be studied after an appropriate
shift, by introducing the symmetry breaking background:
\be
\phi^a\rightarrow (\sqrt{N}\Phi+\phi^1,\phi^i).
\ee
The quantum field representing the fluctuations of the order parameter
is termed $\sigma$, while the modes transversal to it are the Goldstone
pions.
  
The equation of state, which determines the absolute value of the condensate
comes from the requirement of the vanishing quantum expectation for the
coefficient of the term linear in $\phi^1$ in the shifted Lagrangian:
\be
\lag{\delta L\over \delta\phi^1}\rag =0=\sqrt{N}\Phi \Bigl
[m^2+{\lambda \over 6}\Phi^2+{\lambda\over
6N}\lag(\phi^a)^2\rag-{h\over\Phi}\Bigr ].
\label{eqstate}
\ee
The quadratic fluctuations of the shifted fields are computed with an
accuracy ${\cal O}(N^1)$, therefore only the contribution of the pions
is retained.
Anticipating a non-zero Goldstone mass due to the explicit symmetry breaking
(which will be verified below) one finds the following relation 
connecting renormalised quantities (only contributions proportional to $N^0$ 
are kept):
\be
m_R^2+{\lambda_R\over 6}\Phi (T)^2+{\lambda_R\over
  96\pi^2}m_G^2(T)\ln{m_G^2(T)e\over M_0^2}+{\lambda_RT^2\over 12\pi^2}
\int_\frac{m_G(T)}{T}^\infty dy\frac{\sqrt{y^2-\frac{m_G^2(T)}{T^2}}}{e^y-1}=
{h\over\Phi (T)}.
\label{eqstate1}
\ee
Here the following renormalised couplings were introduced (momentum
cut-off was applied for the regularisation of some divergent integrals):
\be
{m^2\over\lambda}+{\Lambda^2\over 96\pi^2}-{m_R^2\over
  96\pi^2}\ln{e\Lambda^2\over M_0^2}=
{m^2_R\over\lambda_R},
\qquad {1\over\lambda} +{1\over 96\pi^2}\ln{e\Lambda^2\over
M_0^2}={1\over\lambda_R}. 
\label{renorm}
\ee
The issue of the normalisation point $M_0$ will be discussed below.

At leading order ($N=\infty$) the only contribution to the self-energy of the
Goldstone modes comes from the tadpole diagram. Therefore one has:
\be
G_G^{-1}(p)=p^2-m^2-{\lambda\over 6}\Phi^2(T)-{\lambda\over
6N}\lag(\phi^a)^2\rag =p^2-{h\over\Phi (T)}.
\label{goldstone-prop}
\ee 
Here the sum representing the mass term was simplified
in view of the equation of state (\ref{eqstate}). This equation implies
that at $N=\infty$ the Goldstone particle is stable and its mass, 
$m_G^2(T)=h/\Phi (T)$ increases with the temperature.

With this identification the renormalised equation of state (\ref{eqstate1})
can be cast into a more practical form by eliminating
$m^2_R$ and $h$ in favour of the $T=0$ value of the condensate ($\Phi_0$) and
of the pion mass ($m_{G0}$):
\bea
&&
{\lambda_R\over 6}{\Phi_0^2\over m_{G0}^2} \left({m_{G0}^4\over m_G^4}-1\right)
+{\lambda_R\over 96\pi^2}\left[\left({m_G^2\over m_{G0}^2}-1\right)
\ln{m_{G0}^2e\over M_0^2}+{m_G^2\over m_{G0}^2}\ln{m_G^2\over m_{G0}^2}\right]
\nonumber\\
&+&{\lambda_R T^2\over 12\pi m_{G0}^2}\int_{m_G/T}^\infty dy
\sqrt{y^2-m_G^2/T^2}(e^y-1)^{-1}={m_G^2\over m_{G0}^2}-1.
\label{mgt}
\eea
This form makes it clear that after $\lambda_R$ and $ M_0/m_{G0}$ are
chosen in the process of 
renormalisation at $T=0$, one finds from this equation $m_G(T)/m_{G0}$
as a function of $T/m_{G0}$ if the physical input $\Phi_0^3/h
\equiv f_\pi^2/Nm_\pi^2$ is made.

The $\sigma$ propagator receives ${\cal O}(N^0)$ contribution from the
infinite iteration of the bubble diagrams $b(p)$, where on both lines  
Goldstone fields are propagating:
\be
G_H^{-1}(p)
=p^2-{h\over\Phi (T)}-{\lambda\over 3}\Phi^2(T){1\over 1-\lambda b(p)/6}.
\label{higgs-prop}
\ee
Without entering its derivation we give here also the expression for the 
leading large $N$ propagator of the quadratic composite field
$(\phi^a\phi^a-\lag(\phi^a)^2\rag)(\x ,t)$:
\be
F(p)={(p^2-h/\Phi (T))b(p)/6+\Phi^2(T)/3\over
(p^2-h/\Phi (T))(1-\lambda_Rb(p)/6)-\lambda_R\Phi^2(T)/3}.
\label{comp-quad}
\ee
It has the same denominator as $G_H(p)$ making explicit the hybridisation
of the two objects.

The bubble contribution with external momentum $p_0,{\bf p}$ is the sum 
of a zero temperature and a $T$-dependent part, $b(p)=b_0(p)+b_T(p_0,{\bf p})$,
which read as follows:
\be
b_0(p)={1\over 16\pi^2}\biggl(-\ln{e\Lambda^2\over M_0^2}+\ln{m_G^2\over 
M_0^2}-\sqrt{1-4m_G^2/p^2}\ln{\sqrt{1-4m_G^2/p^2}-1\over \sqrt{1-4m_G^2/p^2}+1}
\biggr),
\label{bubble0}
\ee
\begin{eqnarray}
  b_T(p) &=& \int\frac{d^3{\bf q}}{(2\pi)^3}
  \,\frac1{4\omega_1\omega_2}\,\left\{  (n_1+n_2)\left[
  \frac{1}{p_0-\omega_1-\omega_2 + i\epsilon} - 
  \frac{1}{p_0+\omega_1+\omega_2 + i\epsilon}\right]\right. \nonumber\\ 
  &&\left.-(n_1-n_2) \left[\frac{1}{p_0-\omega_1+\omega_2 + i\epsilon} - 
  \frac{1} {p_0+\omega_1-\omega_2 + i\epsilon} \right]\right\},
\label{bubbleT}
\end{eqnarray}
where $n_i=1/(\exp(\beta\omega_i)-1)$ and $\omega_1=({\bf q}^2+m_G^2)^{1/2},
\omega_2=(({\bf q}+{\bf p})^2+m_G^2)^{1/2}$. 
The first term in the expression of $b_0(p)$ is cancelled in the expression
of the $\sigma$ propagator by the divergence of the bare coupling $\lambda$
and the inverse propagator, expressed in terms of the renormalised quantities
is finite. Note, that $b_0(p)$ has to be evaluated with $m_G(T)$! 

The phenomenologically most interesting object is the
 spectral function of the order parameter field $\sigma$, defined as
\be
\rho_H (p_0,{\bf p},T)=-{1\over\pi}{\rm Im}G_H(p_0,{\bf p},T).
\ee

\begin{figure}[htbp]
\begin{center}
\includegraphics[width=9cm]{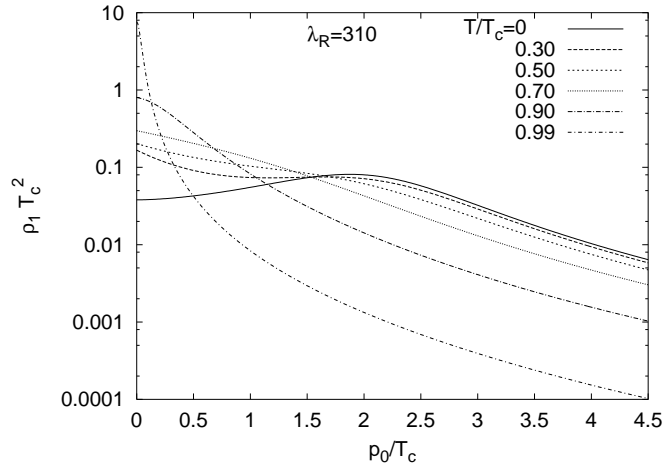}
\end{center}
\caption{Temperature dependence of the modified spectral function 
$\rho_1$ in the $\sigma$ meson channel for $h=0$}
\label{h02point}
\end{figure}

For non-zero $T$ $\rho_H$ has a singularity at $p_0=0$ due to the Bose-Einstein
 factor $n(\beta p_0/2)$. In the chiral limit $h=0$ 
this point coincides with the
two-pion threshold. In order to make the effects related to the physical 
excitations visible around this point
in Fig.\ref{h02point} we show $\rho_1 (p_0,0,T)T_c^2\equiv
(1-\exp (-p_0/2T))\rho_H(p_0,0,T)T_c^2$. The curves are drawn
 for various temperatures below $T_c$ and a representative value of 
$\lambda_R=310$. 

The temperature dependence of the order parameter
needed for its evaluation is obtained from the equation of state which
simplifies in this case \cite{cooper97} to
\be
{\Phi^2(T)\over T_c^2}={1\over 12}\Bigl(1-{T^2\over T_c^2}\Bigr),
\qquad T_c^2=12\Phi_0^2.
\label{chiral-eqstate}
\ee

The shape of the spectral density
 starts with a broad bump at $T=0$ which is shifted towards lower frequencies
 when the temperature is increased. Already for $T=0$
$\rho_H$ has finite value at $p_0=0$, where its value 
is gradually increasing. At a temperature $T\sim 0.7T_c$ the bumpy structure
at finite $p_0$ completely disappears and a simple Lorentzian shape
develops with its maximum located at the threshold.

 This evolution can be transparently interpreted if the temperature
dependent location of the physical pole of the $\sigma$
propagator is found. The physical pole is located in the lower half of the
complex $p_0$-plane, therefore one is faced
with the analytical continuation of $b_T(p)$ into the 
lower $p_0$-halfplane, since originally it is defined only in the upper 
halfplane by the Landau-prescription: $\epsilon >0$. As becomes clear,
this threshold enhancement
is the manifestation of the gradual chiral symmetry restoration as $T$ tends
to $T_c$.

\section{The physical quasiparticle excitation in the $\sigma$ channel}

Below we give some details concerning the temperature variation of the
excitation spectra and the interpretation of the spectral function with 
its help in the chiral limit ($h=0$). 
One starts with the analysis at $T=0$, where one fixes the renormalised
parameters $m_R^2, \lambda_R$, remaining after $h$ was set to zero. 
The renormalised mass is related to the size
 of the condensate (or in view of (\ref{chiral-eqstate}) to $T_c$) 
by the well-known equation: $-m_R^2/\Phi^2_0=\lambda_R/6$. 

$\lambda_R$ will be chosen from a range where one finds for $\Gamma /M_\sigma$
values which are the closest to the experimental one.  For this
we determine the $T=0$ pole of the
$\sigma$ propagator by looking for the zeros of $G_H^{-1}$ in terms of 
renormalised quantities:
\be
G_H^{-1}(p)=p^2-{\lambda_R\over 3}\Phi^2_0{1\over 1-\lambda_R\ln
(-p^2/M_0^2)/96\pi^2}=0.
\label{zeroTeq}
\ee
The physical solution at rest of this equation is parametrised  by putting
${\bf p}=0, p_0=M_0\exp (-i\varphi_0), \pi /2>\varphi_0>0$. 
(One can find its mirror solution in the third quarter). The notation shows, 
that the absolute value of the pole is chosen for the normalisation point 
defined in (\ref{renorm}) and used in (\ref{bubble0}). If a physically
satisfactory
solution is found for some value $\lambda_R$ with this normalisation point, 
the renormalisation group invariance of (\ref{zeroTeq}) ensures that
for a different choice of $M_0$ the same ratio $\Gamma /M_\sigma$ is found
at some appropriately shifted value of $\lambda_R$. In this way 
the actual value of $\lambda_R$ cannot be said to be large or small!

\begin{figure}
\begin{center}
\includegraphics[width=9cm]{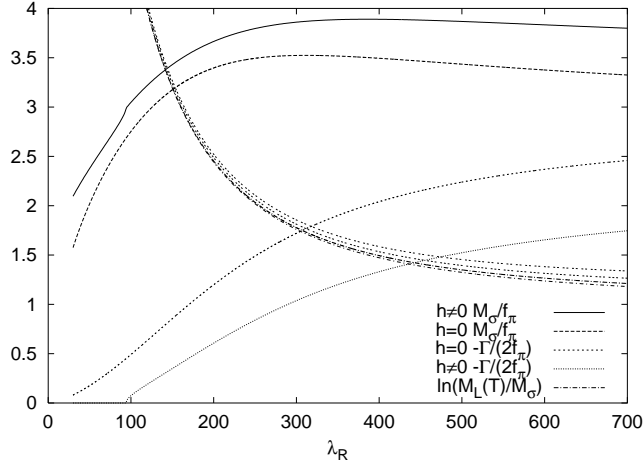}
\end{center}
\caption{The imaginary and real parts of the physical poles at $T=0$ in the
  chiral limit and also for $h\neq 0$. Also shown is the logarithm of the
tachyon pole position in proportion to the mass of $\sigma$ for various 
temperatures below $T_c$ and for $h=0$}
\label{T0-poles}
\end{figure}

 Since in view of
(\ref{eqstate1}) in the chiral limit one has a very simple equation for the
critical temperature, our
solution provides the mass $M_\sigma =M_0\cos\varphi_0$ and the
imaginary part 
$M_0\sin\varphi_0$ in proportion of the critical temperature. 
It is interesting to note that using for $\sqrt{N}\Phi=2\Phi$ the
experimental value of $f_\pi$,
eq.(\ref{chiral-eqstate}) gives $T_c=161$ MeV, while the lattice simulations 
yield $(173\pm 8)$ MeV \cite{karsch01}. The agreement is much better than
expected. The real and imaginary parts of the complex physical pole are
shown in Fig.\ref{T0-poles} as a function of $\lambda_R$.

In addition to the physical zeros of the inverse propagator scalar
theories are known to have a tachyonic zero on the positive imaginary
axis $(p_0=iM_L)$ (see \cite{boyan99} and references therein), which makes the 
theory unstable. Therefore the
theory can be used as an effective theory until the physical scale $M_0$  
is acceptably smaller than $M_L$. The position of the tachyonic zero is also
found from (\ref{zeroTeq}). (The fact that the tachyonic pole shows up
also in $G_H$ not only in $F$ is a manifestation of hybridisation).
It is clear from Fig.\ref{T0-poles} that for
$\lambda_R\leq 400$ one has $M_L/M_0>4$ and the effective approach is
well justified. The representative value $\lambda_R=310$ corresponds
to a ratio $M_\sigma/\Gamma=1$ at $T=0$. The value of $M_\sigma$ is
estimated to be $7\Phi_0\sim 350$ MeV. 

At finite temperature we are interested in the
temperature dependence of the $\sigma$-pole at rest, therefore we set 
${\bf p}=0$ in (\ref{bubbleT}).
The analytic continuation onto the lower plane can be carried out by
adding to the expression of $b_T$ valid in the upper halfplane a term
 which makes its imaginary part
continuous when one approaches the real $p_0$-axis either from the
upper or the lower halfplane:
\bea
b_T^{>}(p_0)&=&{1\over 8\pi^2}\int_0^\infty dx{1\over e^x-1}\Bigl[{1\over
z-x}-{1\over z+x}\Bigr], \qquad z={p_0\over 2T}, \qquad {\rm Im}~p_0>0, 
\nonumber\\
b_T^{<}(p_0)&=&b_T^{>}(p_0)-{i\over 4\pi}{1\over \exp
(z)-1},\qquad {\rm Im}~p_0<0,\qquad {\rm Re}~p_0>0 .
\label{bT-up-down}
\eea

The roots are parametrised the same way as in the $T=0$
case, $p_0=M(T)\exp (-i\varphi)$. After one determines   $\Phi (T)$ from
Eq.(\ref{chiral-eqstate}), from the complex equation
\bea
&&
1-{\lambda_R\over 96\pi^2}\bigg(\ln{M^2(T)\over M^2_0}-i(2\varphi +\pi
)\bigg)-{\lambda_R \over 3}{\Phi^2(T)\over M^2(T)}e^{2i\varphi }
-{\lambda_R\over 6}b_T^>(p_0=Me^{-i\varphi})
\nonumber\\
&+&i{\lambda_R\over 24\pi
}{1\over \exp (M\exp (-i\varphi )/2T) -1}=0
\label{compleq}
\eea
one can find the complex pole as a function of the temperature.
All quantities are measured in units of $T_c\sim\Phi_0$.
 
Also at finite temperature one can study the tachyon  
solution in the upper halfplane using $b_T^>(iM_L)$. One might wonder if the 
temperature dependence of the tachyonic pole does not
restrict further the coupling range where the effective use of the
scalar theory is consistent. The broadening in Fig.\ref{T0-poles}
 of the line
corresponding to the tachyonic pole reflects the slight decrease of its
mass scale with increasing temperature. However, it is clear that in the 
region $\lambda_R<400$, where the effective approach is consistent for
$T=0$, the position of the tachyon pole is practically unchanged.

Similarly, one can look directly for physical roots along the negative 
imaginary axis. Then using $b_T^<$ from (\ref{bT-up-down}) one finds that 
above a well-defined $T_{imag}(\lambda_R)<T_c$ the physical pole becomes 
purely imaginary: $-iM(T)$. If $\mu\equiv M(T)/2T<<1$, for the highest 
occupied low frequency region, one can use the expansion of
 the Bose-Einstein factors appearing in the last two terms of (\ref{compleq})
with respect to the powers of $\mu$ and keep the leading terms:
\be
1-{\lambda_R\over 48\pi^2}\ln {2T\mu\over M_0}+{\lambda_R\over
3}{\Phi^2(T) \over 4T^2}{1\over \mu^2}-{\lambda_R\over 48\pi}{1\over \mu}=0.
\ee
This equation could have been derived directly, if one would have
applied from the beginning the same approximation in Eq.(\ref{bubbleT}).
Clearly, the region around the origin (the critical region) can be analysed
also directly with help of this simpler equation. Note that one has to go one
step beyond the classical approximation to achieve a consistent approach.

The general pattern of the trajectory of the physical pole with increasing
temperature was the following. The real part of its position started to
diminish when $T$ was raised. In this 
way the broad bump of the spectral function moves towards the origin, and
its width is increasing due to the slight increase of its imaginary part. 
Depending
on $\lambda_R$ the root reaches at some $T_{imag}<T_c$ the negative 
imaginary axis and collides with its mirror root arriving from the third 
quarter. Each one is joining smoothly the trajectory of one of the
pair of imaginary
roots which appear just at $T_{imag}$ and move opposite directions. 
We observe that the $\sigma$-bump gets lost in the background before
the temperature reaches $T_{imag}$. 
The root approaching the origin along the imaginary axis
produces in the spectral function a shrinking shape which becomes a
Lorentzian only very close to $T_c$. Eventually for $T=T_c$ it builds up 
a term proportional to $\delta (p_0)/p_0$.
 
We see that the complex pole of the 
$\sigma$-propagator qualitatively accounts for the behaviour of the
spectral function near and below the critical point as well.
Note that the other pole moving away from the origin along the imaginary axis
as well as the tachyon disappear from $G_H$ at $T_c$. They remain poles of $F$ 
only.

In the vicinity of the critical point, where the $\xi =T/8\pi\Phi^2(T)$
is the dominant length scale and the condition $p_0,|{\bf p}|<<T$ is 
fulfilled, one can derive an equation also for the soft modes with nonzero
momentum:
\be
-3i|{\bf p}|\xi {p_0^2-{\bf p}^2\over |{\bf p}|^2}\ln{p_0-|{\bf p}|
\over p_0+|{\bf p}|}
-{1\over 4\pi}{(p_0^2-{\bf p}^2)\xi \over T_c}\ln{p_0^2-{\bf p}^2\over T_c^2}
=1.
\ee
Its solution in the approximation, when on the left hand side only the
first term is retained exhibits the form of dynamical scaling \cite{ferrell67,
hohenberg77}: $p_0=|{\bf p}|^{\tilde z} f(|{\bf p}|\xi)$ with 
$\tilde z=1$.
The second term provides the leading correction to scaling. In $O(N)$ models,
the dynamical exponent $z=d/2$ has been obtained for finite $N$ on the basis 
of scaling and renormalisation group arguments, where in our case $d=3$
\cite{sasvari75,sasvari74,rajagopal93}. For the correct interpretation of the
situation it is important that there are two distinct hydrodynamical regions
in the $O(N)$ model for large $N$ \cite{sasvari75}. (The system studied in
\cite{sasvari75} can be regarded as a lattice regularised version of the
linear $\sigma$-model). In the true critical 
region $z=d/2$ is valid, in a precritical region $\tilde z=1-8S_d/Nd+
{\cal O}(1/N^2)$, where $S_d=2/\pi^2$ for $d=3$. The first region shrinks
when $N$ becomes large and completely disappears at $N=\infty$. Furthermore,
it has been found that $\tilde z$ agrees with the dynamical exponent 
determined for the quasiparticles, that is outside the hydrodynamical regime,
at least to ${\cal O}(1/N)$\cite{szepfalusy74}. Details of the application 
of this analysis to the present case will be published in our forthcoming
paper.

\section{The case of the explicit breaking of chiral symmetry}

Finally, we wish to discuss shortly the results of the application of
the leading large $N$ solution of the $O(N)$ model to the low lying
effective  $\pi -\sigma$ system 
where an explicit symmetry breaking is necessary.

Since in this case one cannot speak of any phase transition,
it is more sensible to look for the roots of $G_H^{-1}(p_0,0)$ in proportion to
$\Phi_0\equiv f_\pi /2$. The value $m^2_{G0}/\Phi_0^2\approx
9.06$ is fixed by phenomenology and 
choosing $\lambda_R$ with the restriction $M_0/M_L\leq 4$, one can find the
zero temperature position of the $\sigma$-pole. In 
Fig.\ref{T0-poles} its real and imaginary parts are already 
shown as a function of $\lambda_R$ together with the results obtained for 
$h=0$. In the region of $\lambda_R$ allowed the ratio
$M_\sigma/\Gamma_\sigma$ moves away from the phenomenologically preferred
range ($M_\sigma=3.95 f_\pi, M_\sigma/\Gamma\sim 1.4$) 
emphasising the need for a next-to-leading order calculation.

\begin{figure}
\begin{center}
\includegraphics[width=7.5cm]{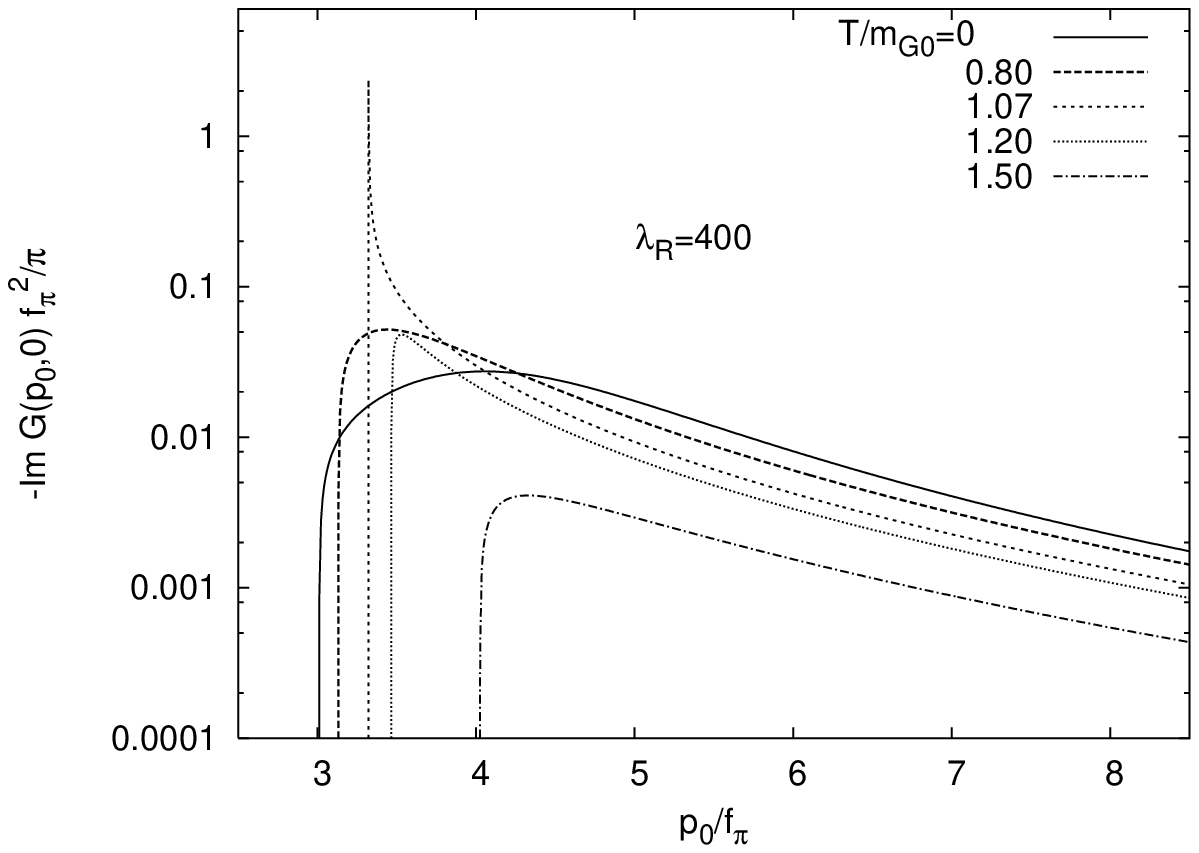}
\includegraphics[width=7.5cm]{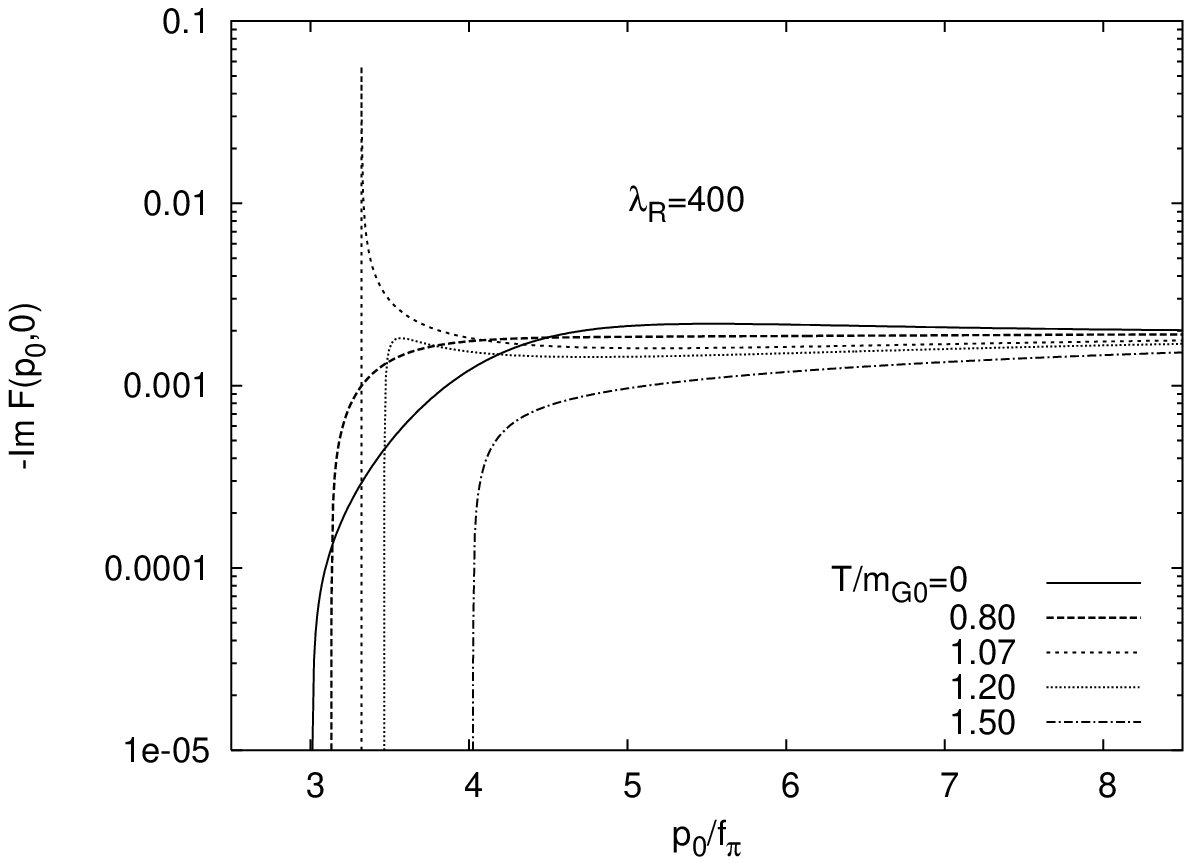}
\end{center}
\caption{The spectral functions for the linear sigma model with
  realistic couplings and explicit symmetry breaking. On the left
  figure the imaginary part in the $\sigma$-channel, on the right the
  same for the composite channel is shown}
\label{h-spectral}
\end{figure}

Now $\Phi (T)$ is determined from (\ref{mgt}).
The spectral function of the single $\sigma$-channel
is shown in the left of Fig.\ref{h-spectral} for various values of the
temperature 
in the transition range $T=(0-1.2)m_{G0}$ with $\lambda_R=400$.
 The location of its
maximum at $T=0$ is correlated with the location of the $T=0$ pole. 
Maximal threshold enhancement \cite{hatsuda87} is observed in the close
neighbourhood of $1.074m_{G0}$ with the spectral function numerically 
exhibiting a cuspy form, with a very narrow
width. However, when the temperature is further increased, the
maximum is pulled away from the actual position of the threshold and a
broad rounded structure is seen. We return to the
detailed interpretation of this structure in terms of the pole
trajectories in a forthcoming publication.
     
As we have mentioned above also the spectral function
belonging to the propagator (\ref{comp-quad})
of the composite field  can be constructed to leading order in $N$.
The corresponding spectral function reflects the excitations in the composite
$(\phi^a)^2$ channel. Its temperature dependent deformation is displayed on
the right hand of Fig.\ref{h-spectral} for 
the same (realistic) parameters as used in case of $\rho_H$ above. 
Its behaviour is rather similar to the single meson spectral
function. No signal of any meson-meson bound state can be observed.
The only difference one notices is that for large frequencies this
function approaches a constant, which reflects mostly the large 
residuum of the tachyon pole in this propagator.

\section{Conclusions}
We have described the temperature dependence of the elementary
excitations of the $O(4)$ model in the leading order of the $1/N$
expansion. For the chirally symmetric case a very suggestive picture
of the complex pole evolution makes unique the interpretation of the
change of shape of the single-particle spectral function when the
temperature approaches the critical point. 

The restrictions on the range of the scalar self-coupling $\lambda_R$
arising from the requirement of keeping distance from the tachyonic pole,
for a fixed normalisation point $M_0$
prevent us in finding a fully realistic $\sigma$-particle mass at $T=0$
when the physical pion mass is the input. We have found that the
threshold enhancement in both the elementary and the composite spectral 
functions is maximal at some $T_{enh}<m_{G0}$.  
Beyond this temperature the cuspy maximum becomes rounded again. We believe
that this qualitative
feature remains valid when the next to leading order corrections will be
included.

\section*{Acknowledgement}

This research was supported by the Hungarian Research Fund (OTKA).


\begin{thebibliography}{9}
\bibitem{hatsuda01} for a recent review see, T. Hatsuda,
Nucl. Phys. {\bf A698} (2002) 243
\bibitem{chiku98}S. Chiku and T. Hatsuda, Phys. Rev. {\bf 58} (1998)
  076001
\bibitem{kondor74}P. Sz{\'e}pfalusy and I. Kondor, Ann. Phys. (N.Y.), {\bf 82} 
(1974) 1, L. Sasv\'ari and P. Sz{\'e}pfalusy, J. Phys. {\bf C7} (1974) 1061
\bibitem{griffin93}A. Griffin, Excitation in a Bose-Condensed Liquid, Cambridge
University Press, Cambridge, 1993
\bibitem{szepfalusy76}P. Sz{\'e}pfalusy, in {\it Critical Phenomena}, 
Lecture Notes in Physics {\bf 54}, eds. J. Bray and R.B. Jones, 
Springer 1976, p.112 
\bibitem{cooper97}F. Cooper, S.Habib, Y. Kluger and E. Mottola, 
Phys. Rev. {\bf 55} (1997) 6471 
\bibitem{cooper94} F. Cooper, S. Habib, Y. Kluger, E. Mottola, J.P. Paz
  and P.R. Anderson, Phys. Rev. {\bf D50} (1994) 2848
\bibitem{boyan99}D. Boyanovsky, H.J. de Vega, R. Holman and J. Salgado,
  Phys. Rev. {\bf D59} (1999) 125009
\bibitem{baacke00} J. Baacke and K. Heitmann, Phys. Rev. {\bf D62}
  (2000)  105022
\bibitem{berges01} J. Berges, Nucl. Phys. {\bf A699} (2002) 847
\bibitem{aarts02} G. Aarts, D. Ahrensmeier, R. Baier, J. Berges and
  J. Serreau, hep-ph/0201308
\bibitem{tornquist02} N.A. T{\"o}rnquist, hep-ph/0201171
\bibitem{karsch01}F. Karsch, E. Laermann and A. Peikert,
  Nucl. Phys. {\bf B605} (2001) 579
\bibitem{ferrell67} R.A. Ferrell, N. Menyh\'ard, H. Schmidt, F. Schwabl and
P. Sz{\'e}pfalusy, Phys. Rev. Lett. {\bf 18} (1967) 891,
B. I. Halperin and P.C. Hohenberg, Phys. Rev. Lett. {\bf 19} (1967) 700
\bibitem{hohenberg77} P.C. Hohenberg and B.I. Halperin, Rev. Mod. Phys. 
{\bf 49} (1977) 435
\bibitem{sasvari75} L. Sasv\'ari, F. Schwabl and P. Sz{\'e}pfalusy, Physica 
{\bf 81A} (1975) 108
\bibitem{sasvari74} L. Sasv\'ari and P. Sz{\'e}pfalusy, Physica {\bf 87A} 
(1977) 1
\bibitem{rajagopal93}K. Rajagopal and F. Wilczek, Nucl. Phys. {\bf 393} (1993)
\bibitem{szepfalusy74} P. Sz{\'e}pfalusy and L. Sasv\'ari, Acta Phys. Hung.
{\bf 37} (1974) 343
\bibitem{hatsuda87}T. Hatsuda and T. Kunihiro, Phys. Lett. {\bf B185} 
(1987) 304
\end{thebibliography}
\end{document}